\documentclass[a4,12pt]{article}
\usepackage{amsmath,amsfonts,amssymb,graphicx,epsfig}
\usepackage{hyperref}
\usepackage{indentfirst}
\begin{document}
\title{\bf On topological spin excitations on a rigid torus}
\author{V.L. Carvalho-Santos $^{a, b}$, A. R. Moura$^{a}$,
W.A. Moura-Melo$^{a}$\footnote{E-mail address: winder@ufv.br} , and A. R. Pereira$^{a}$
\\
\it\small $^{a}$ Departamento de F\'isica, Universidade Federal de
Vi\c cosa\\\it\small 36570-000, Vi\c cosa, Minas Gerais, Brazil\\
\it\small $^{b}$ Escola Agrot\'ecnica Federal de Senhor do
Bonfim\\\it\small 48970-000, Senhor do Bonfim, Bahia, Brazil\\}
\date{}
\maketitle
\begin{abstract}
We study Heisenberg model of classical spins lying on the toroidal support, whose internal and external radii are $r$ and $R$, respectively. The isotropic regime is characterized by a fractional soliton solution. Whenever the torus size is very large, $R\to\infty$, its charge equals unity and the soliton effectively lies on an infinite cylinder. However, for $R=0$ the spherical geometry is recovered and we obtain that configuration and energy of a soliton lying on a sphere. Vortex-like configurations are also supported: in a ring torus ($R>r$) such excitations present no core where energy could blow up. At the limit $R\to\infty$ we are effectively describing it on an infinite cylinder, where the spins appear to be practically parallel to each other, yielding no net energy. On the other hand, in a horn torus ($R=r$) a singular core takes place, while for $R<r$ (spindle torus) two such singularities appear. If $R$ is further diminished until vanish we recover vortex configuration on a sphere. 
\end{abstract}

\section{Introduction and motivation}
Geometrical and topological concepts and tools are important in many branches of natural sciences, namely in Physics. For instance, the idea of symmetry, intimately associated to geometry, is a keystone for studying a number of fundamental properties of several physical systems, e.g., Noether theorem asserts a conserved quantity to each continuous symmetry of the associated action. Topology, in turn, is crucial for classifying and for giving stability to certain excitations, like solitons (extending objects having finite energy) and vortices (presenting non-vanishing vorticity around a given singular point or a topological obstruction). These and others have been observed in a number of systems, like superconductors, superfluids, and magnetic materials, namely vortex-like magnetization has been directly observed in nanomagnets \cite{ShinjoWacho}. On the other hand, the observed vortex-pair dissociation is the mechanism behind the topological phase transition\cite{BKT}. Another kind of topological objects emerges from Euclidian non-Abelian pure-field models in (3+1) dimensions. They are called instantons, since they represent points in the Euclidian space-time. In this case, the relevant homotopy group is that associated to the mapping of the field space to the 4-dimensional Euclidian one, so that $\pi_{3}(S^{3} \to S^{3})={\bf Z}$, for example, if the internal space is the QCD gauge group, $SU(3)$. Such excitations play important role in this framework once the nonconservation of the flavor singlet axial current is proportional to the instanton charge \cite{QCD}.\\

In turn, the toroidal geometry has recently received a considerable attention, for example, as tight traps for Bose-Einstein condensates, inside which the atoms develop quasi-one-dimensional confined dynamics subject to periodic boundary conditions\cite{BECtoro}. Ring-shaped carbon nanotubes have been shown to provide quasi-zero-dimensional systems, whenever the rings are very small\cite{ringCNT,ZhaoPLA}. In addition, their topology also make them possible field-effect transistors for technological applications\cite{WatanabeCNT}, as well as, the suitable control of their stable magnetization (e.g., those with vortex-like profile) may be useful for applications in magneto-electronic devices, namely if an array of magnetic nanorings is concerned \cite{ShekaVavassori}. From a more fundamental point of view, it has been verified that whenever the Ising model is defined on a donut-shaped lattice, then instead of an unique, there appear two critical temperatures\cite{ShimatoroJMMM2007}. Even in Biology, the toroidal shape plays important roles: for instance, it has been observed that a large number of proteins involved in DNA metabolism adopts a ring-like shape, even though these proteins have quite distinct and unrelated functions in this mechanism. Why this geometry/topology is so abundant in this process and the reason why life evolution has selected it amongst many others remains a puzzle in the specialized literature \cite{toroBio}.\\

Geometrically, a torus of genus-1 (one central hole) is a compact surface whose (Gaussian) curvature smoothly varies from $-1/r(R-r)$ to $+1/r(R+r)$, along its polar angle (see details below), so interpolating between the pseudospherical and spherical curvatures whenever $R>r$. From the topological point of view, the simplest genus-1 torus is obtained by the topological product of two circles, $T^1=S^1 \wedge S^1$ \cite{Nakahara}. [In general, a genus-n torus is got in a similar way, for example, that of genus-2 is given by $T^2=T^1\wedge T^1$]. This confers it a non-simply connectivity, say, closed loops holding on its central hole or the polar circumference cannot be shrunk to a point (other non-simply connected topologies are the ordinary circle and a disk with a hole, an annulus).\\

For studying classical spin-like textures, of solitonic and vortex kinds, on a torus, we first write down the continuum version of the Heisenberg (exchange) Hamiltonian in this geometry (Section 2). Later, the non-linear Euler-Lagrange equations are obtained and some particular cases are explicitly considered for describing the desired excitations. In addition, their profiles, energies and other basic properties are discussed and compared to their counterparts from other surfaces. These tasks are performed in Section 3, where solitonic-like solutions in the isotropic regime are described and, in Section 4, which is dedicated to vortex-like configurations, studied within the Planar Rotator Model (equivalently, the XY model, once are dealing only with static properties). Finally, we close our paper by pointing out our Conclusions and Prospects for forthcoming investigation.\\

\section{The continuum Heisenberg model on the torus}
The anisotropic exchange Heisenberg model, for nearest-neighbor interacting
spins on a two-dimensional lattice, is given by the Hamiltonian
below:
\begin{equation}
\label{heisdisc} H_{latt}=-J'\sum_{<i,j>}{{\cal H}_{i,j}}=-J'\sum_{<i,j>}
(S_{i}^{x}S_{j}^{x}+S_{i}^{y}S_{j}^{y}+(1+\lambda)
S_{i}^{z}S_{j}^{z})
\end{equation}
where $J'$ denotes the coupling between neighboring spins, and
according to $J'<0$ or $J'>0$, the Hamiltonian describes a ferro or
anti ferromagnetic system, respectively.
$\vec{S_{i}}=(S_{i}^{x},S_{i}^{y},S_{i}^{z})$ is the spin operator
at site $i$, the parameter $\lambda$ accounts for the anisotropy
interaction amongst spins: for $\lambda>0$, spins tend to align
along the (internal) Z-axis (easy-axis regime); for$\lambda=0$, one
gets the isotropic case; for $-1<\lambda<0$, we have the easy-plane
regime, while the $\lambda=-1$ case, yields to the so called XY model
(or the Planar Rotator Model, PRM, if we focus on 2-component spin,
imposing $S_{z}\equiv0$, so that $\vec{S}_{\rm PRM}=(S_{x}, S_{y})$).

In the continuum approach of spatial and spin variables, valid at
sufficiently large wavelength and low temperature, the Hamiltonian (\ref{heisdisc}) may be written as $(J\equiv J'/2)$ \cite{SaxenaPhysA,nossapseudoPLA,nossaesferaPLA}:
$$
\label{heiscont} H_{1}=J\int\int\sum_{i,j=1}^{2}\sum_{a,b=1}^{3}
g^{ij}h_{ab}(1+\delta_{a3}\lambda)\biggl(\frac{\partial
S^{a}}{\partial\eta_{i}}\biggl)\biggl(\frac{\partial
S^{b}}{\partial\eta_{j}}\biggl)\sqrt{|g|}d\eta_{1}d\eta_{2}
=$$\begin{equation}
=J\int_\Omega\int(1+\delta_{a3}\lambda)(\vec{D}S^{a})^{2}d\Omega
\end{equation} where $\Omega$ is the surface with curvilinear
coordinates $\eta_{1}$ and $\eta_{2}$, so that
$d\Omega=\sqrt{|g|}d\eta_{1}d\eta_{2}$, $\delta_{a3}$ is the
Kronecker symbol, $\vec{D}$ is the covariant derivative, $\sqrt{|g|}=\sqrt{|{\rm det}[g_{ij}]|}$,
$g_{ij}$ and $h_{ab}$ are the elements of surface and spin space metrics, respectively (as usual, $g_{ij}g^{jk}=\delta^k_{\;i}$). Now,
$\vec{S}=(S_{x},S_{y},S_{z})\equiv(\sin\Theta\cos\Phi,\sin\Theta\sin\Phi,\cos\Theta)$
is the classical spin vector field valued on a unity sphere
(internal space), so that $\Theta=\Theta(\eta_{1},\eta_{2})$ and
$\Phi=\Phi(\eta_{1},\eta_{2})$ . With this Cartesian parametrization for $\vec{S}$ we have $h_{ab}=\delta_{ab}$. The Hamiltonian (\ref{heiscont}) may
be also viewed as an anisotropic non-linear $\sigma$ model
(NL$\sigma$M), lying on an arbitrary two dimensional geometry, so
that our considerations could have some relevance to other branches
like hydrodynamic, superfluidity and superconductivity.\\

\begin{figure}
\begin{center}
\includegraphics[width=5.5cm]{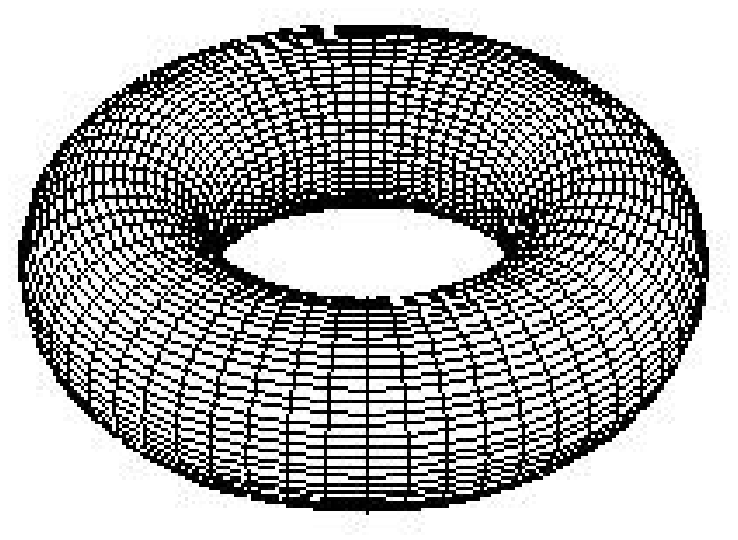}\hskip .2cm
\includegraphics[width=9cm]{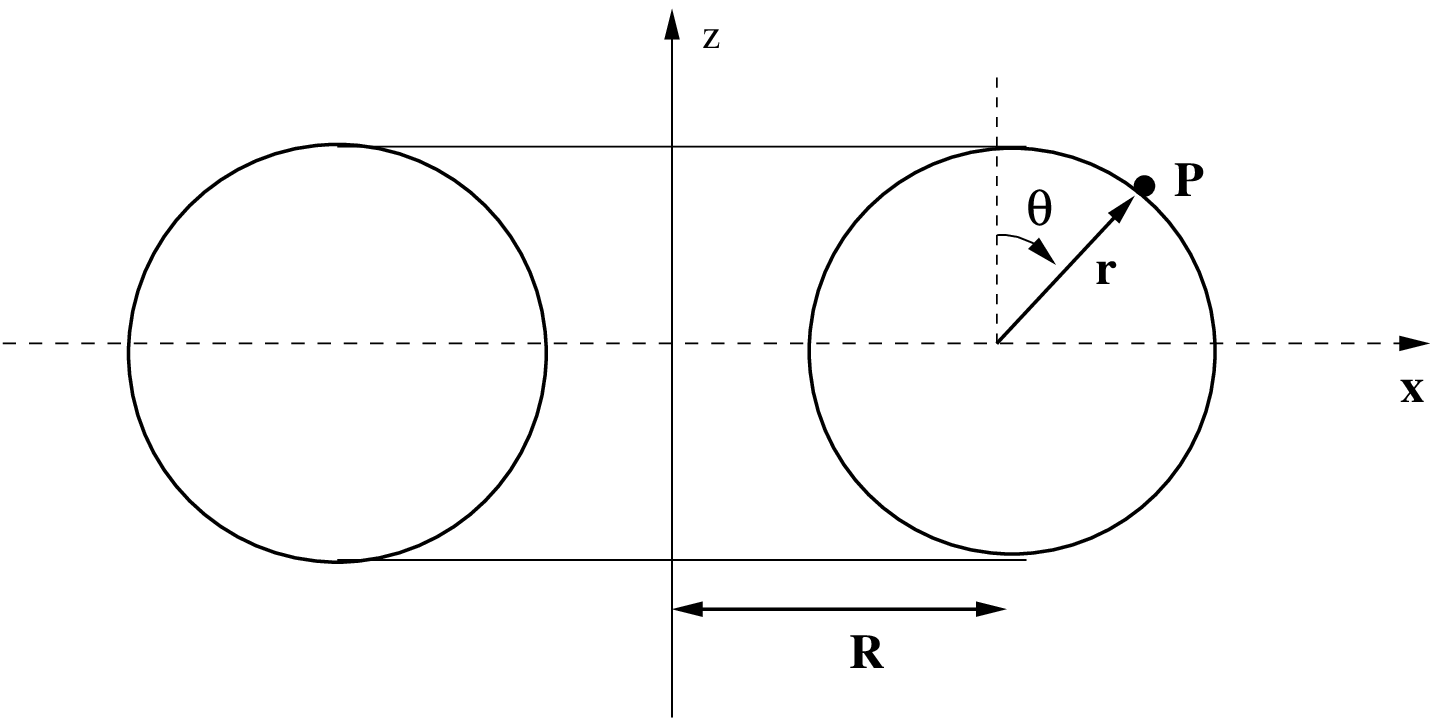} \caption{\small Shape of an
ordinary ring torus embedded in a three-dimensional space (Left). Its cross section and the variables used throughout this work (Right). Azimuth-like angle, $\phi$, runs along the torus tube, $z$-constant planes, and is not shown in the cut.} \label{ring-torus}
\end{center}
\end{figure}

\begin{figure}[!htb]
\begin{center}
   { \includegraphics[width=5cm]{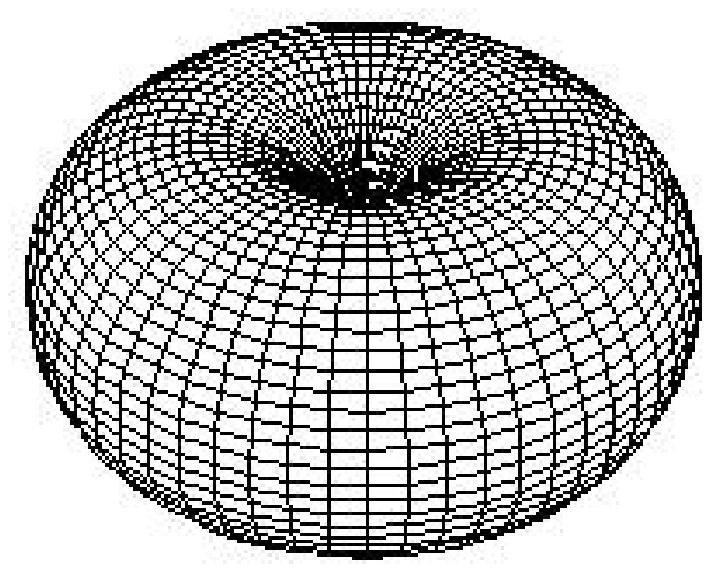}  \hskip .5cm                             \includegraphics[width=7cm]{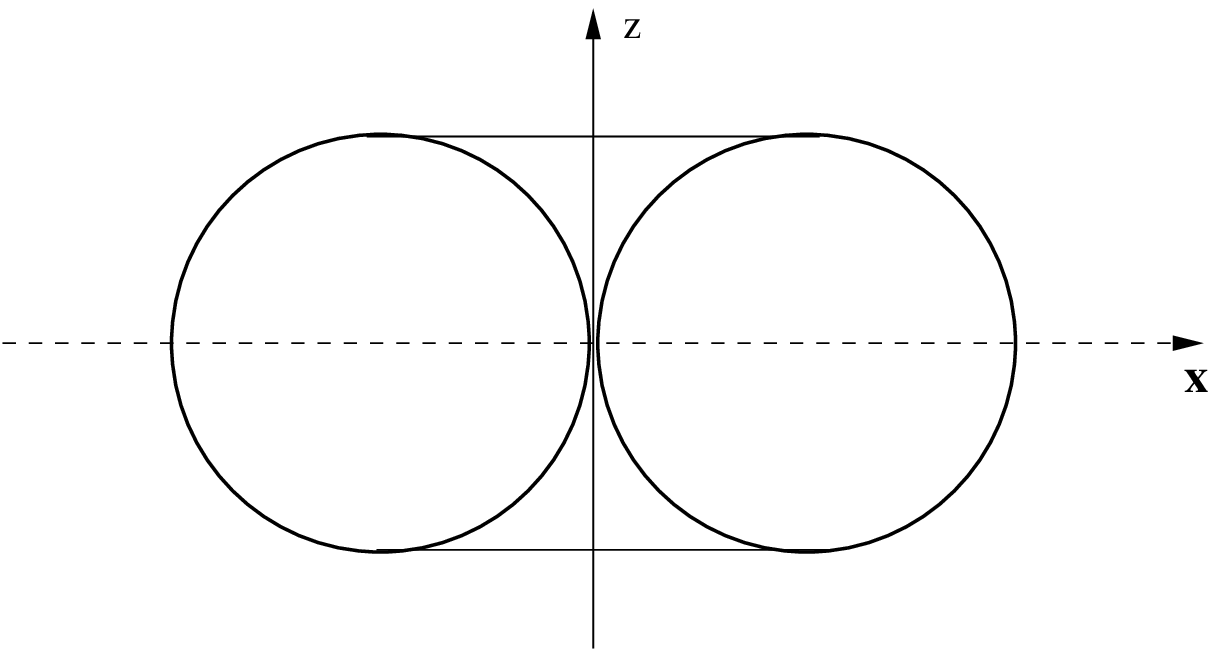}   } \vskip.2cm \hskip -.7cm
   {  \includegraphics[width=5cm]{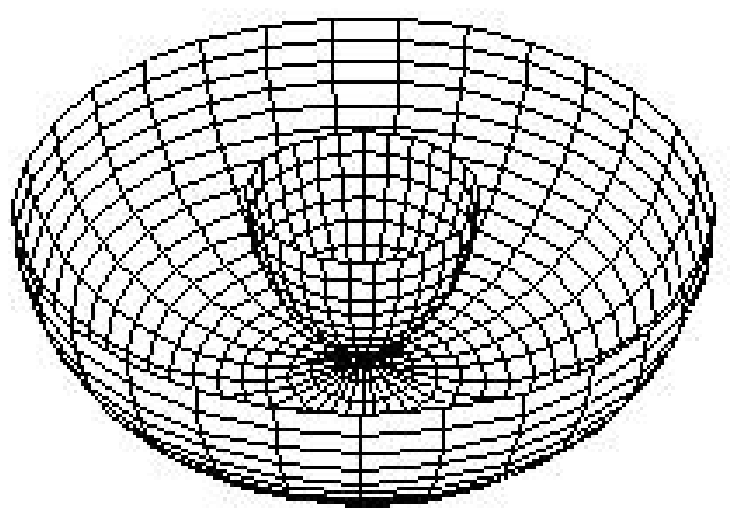} \hskip .5cm             \includegraphics[width=5.6cm]{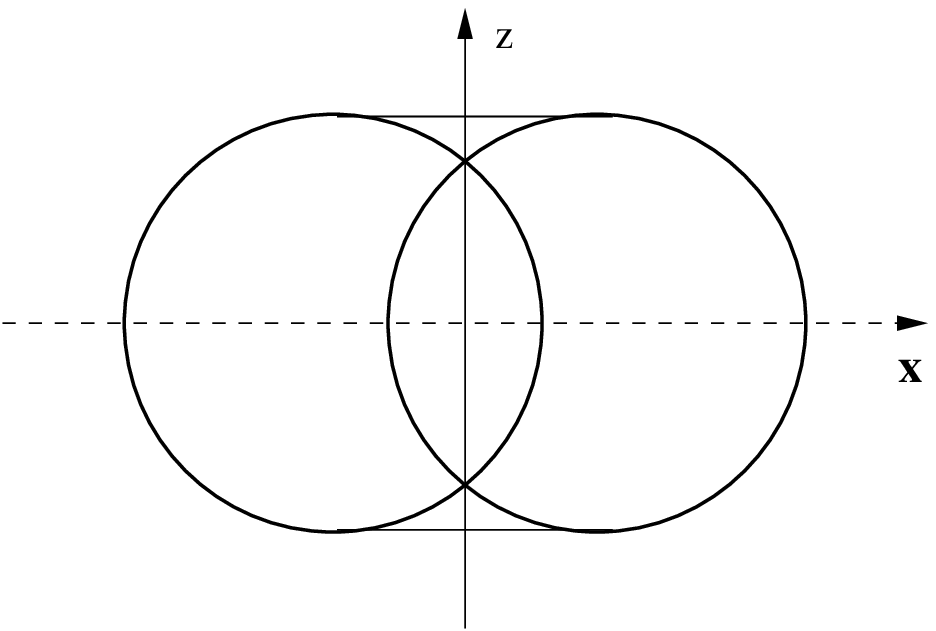}       }
\caption{\small Global and cross-section views of a horn (upper) and spindle tori (in this case, the torus was cut for improving global view). The parameters are those from Fig. \ref{ring-torus}. Note that in the cross-sections the horn and the spindle tori have a single and a pair of self-intercepting points, respectively. Indeed, in the case of the spindle torus such points correspond to two circumferences along which this surface crosses itself.}
\label{horn-and-spindle}
\end{center}
\end{figure}

Our interest is to study the model above on the torus geometry which
is a smooth surface with varying curvature. The simplest torus is thought as a surface having genus
one (a single central hole), and whenever embedded in three
dimensional space it shapes like a donut (see Fig. \ref{ring-torus}). The standard tori are classified in three distinct types concerning the relations between their internal, $r$, and external, $R$, radii. For $R>r$, we have a {\em ring torus} (donut shape, as shown in Fig. \ref{ring-torus}), if $R=r$ a {\em horn torus} is obtained, while for $R<r$ a {\em spindle torus} is described (the latter ones are depicted in Fig. \ref{horn-and-spindle}).\\

Any ordinary torus may be parametrized in a simple way by some distinct
coordinate systems, like the Cartesian \cite{mathem} and the peripolar \cite{geotorus} $(\theta,\phi)$, so that:
\begin{equation}
\left(R-\sqrt{x^{2}+y^{2}}\right)^{2}+z^{2}=r^{2}\,,
\end{equation}
with the parametric equations
\begin{equation} \label{coord1}
x=(R+r\sin\theta)\cos\phi\,,\hspace{0.6cm}y=(R+r\sin\theta)\sin\phi \,,\hspace{0.6cm}
 z=r\cos\theta
\end{equation}
where $R$ and $r$ are the rotating (external) and axial (internal) radii, respectively (see Fig. \ref{ring-torus}). In peripolar variables $(\theta, \phi)$ the metric elements read:
\begin{equation}
\label{met1}
g_{\theta\theta}=r^2\,,\quad g_{\phi\phi}=(R+r\sin\theta)^2\,,\quad 
{\mbox{\rm and}} \quad g_{\phi\theta}=g_{\theta\phi}=0, 
\end{equation}
from what there follows the Gaussian curvature:
\begin{equation}
K=\frac{\sin\theta}{r(R+r\sin\theta)}\,.\label{curvature}
\end{equation}
Note that $K$ varies from $-1/r(R-r)$, at $\theta=3\pi/2$, to $+1/r(R+r)$, at $\theta=\pi/2$. At $\theta=0,\pi$ the curvature vanishes. Indeed, as $R\to0$ we get the sphere geometry, once that $K(R=0)=r^{-2}$, while for $R\to\infty$ we obtain a curvatureless infinite surface, like an infinite cylinder or an annulus. These points will be important later, whenever discussing upon the behavior of solitonic and vortex energies at these limits. Although we shall treat explicitly the case of a ring torus, $0<r<R$, our approach may be extended to the other standard tori, unless otherwise specified.\\

In this set of coordinates the Hamiltonian (\ref{heiscont}) gets the form below ($\partial_\phi\equiv \frac{\partial}{\partial\phi}$ and $\partial_\theta\equiv \frac{\partial}{\partial\theta}$): 
$$
H=J\int_{-\pi}^{\pi}\int_{0}^{2\pi}\Biggr\{\frac{r}{R+r\sin\theta}\left[ \left(1+\lambda\sin
^{2}\Theta\right)\left({\partial_\phi\Theta}
\right)^{2}+\sin
^{2}\Theta\left({\partial_\phi\Phi}\right)^{2}\right]+$$
\begin{equation}+\frac{R+r\sin\theta}{r}\left[\left(1+\lambda\sin^{2} \Theta\right)\left({\partial_\theta\Theta}
\right)^{2}+\sin
^{2}\Theta\left({\partial_\theta\Phi}\right)^{2}\right]\Biggr\}d\phi
d\theta\,,\label{Htoro}
\end{equation}

from what there follow the (static) Euler-Lagrange
equations for $\Theta$ and $\Phi$, respectively:
$$\sin\Theta\cos\Theta\Biggr\{\frac{r}{R+r\sin\theta}\left[\lambda\left(\partial_{\phi}\Theta\right)^{2}+
\left(\partial_{\phi}\Phi\right)^{2}\right]+\frac{R+r\sin\theta}{r}\left[\lambda\left(\partial_{\theta}\Theta\right)^{2}+
\left(\partial_{\theta}\Phi\right)^{2}\right]\Biggr\}=$$$$\cos\theta\left(1+\lambda\sin^{2}\Theta\right)
\left(\partial_{\theta}\Theta\right)+\frac{R+r\sin\theta}{r}\left[(1+\lambda\sin^{2}\Theta)
\partial^{2}_{\theta}\Theta+2\lambda\sin\Theta\cos\Theta\left(\partial_{\theta}\Theta\right)\right]+$$
\begin{equation}
\label{TH}
+\frac{r}{R+r\sin\theta}\left[(1+\lambda\sin^{2}\Theta)\partial^{2}_{\phi}\Theta
+2\lambda\sin\Theta\cos\Theta\left(\partial_{\phi}\Theta\right)\right]\end{equation}\vspace{0.3cm}
\begin{equation}
\label{PH} \cos\theta\sin^{2}\Theta\partial_{\theta}\Phi+
\frac{R+r\sin\theta}{r}\partial_{\theta}\left(\sin^{2}\Theta\partial_{\theta}\Phi\right)
+\frac{r}{R+r\sin\theta}\partial_{\phi}\left(\sin^{2}\Theta\partial_{\phi}\Phi\right)=0
\end{equation}

As expected, the general anisotropic regime of the Heisenberg model
is described by non-linear differential equations. Suitable
non-trivial solutions can be obtained provided that some conditions are
imposed, so that special solutions may be explicitly worked out. At this point, we should note that equations above resemble in form those
counterparts for the planar, spherical and pseudospherical surfaces. Indeed, whenever $R+r\sin\theta$ is identified
with $\mathfrak{r}$, $\mathfrak{R}\sin\theta$ or
$\mathfrak{\varrho}\tau$, while $\phi$ keeps its role as the
azimuth-like angle, expression above exactly recover their planar, spherical
or pseudospherical analogues \cite{nossapseudoPLA,nossaesferaPLA}. Above, $\mathfrak{r}=\mathfrak{|\vec{r}|}$ is the planar radial distance,
$\mathfrak{R}$ is the sphere radius while
$\mathfrak{\varrho}\tau$ accounts for the distance measured along a
pseudospherical geodesic, say, a hyperbole.

\section{The isotropic model and solitonic solutions}

\begin{figure}[!h]
\begin{center}
   { \includegraphics[width=10cm]{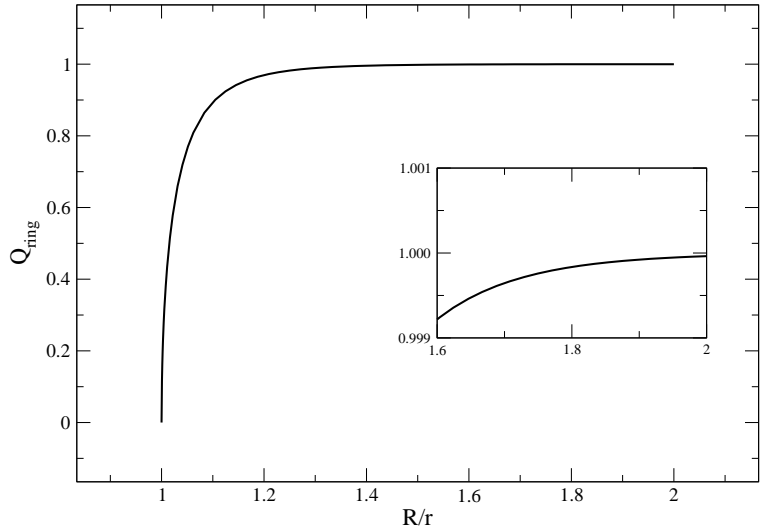}  
   }
\caption{\small How the solitonic charge associated to solution (\ref{solitontoro}) behaves in the ring torus as function of $R/r>1$. Although $Q$ closes the unity very fast (e.g., for $R/r=2$ we obtain $Q_{R=2r}\approx 0.99995$; see inset plot) it should be stressed that only when $R\to\infty$ we get $Q=1$.}
\label{Q-ring}
\end{center}
\end{figure}

The simplest way to seek for possible solitonic solutions associated to the present model on a torus is considering the isotropic regime, $\lambda=0$, and writing down the Hamiltonian (\ref{Htoro}), and its associated equations (\ref{TH},\ref{PH}), in a more suitable coordinate system which allows us of getting the sine-Gordon equation in a simpler way (further details may be found in Refs.\cite{SaxenaPhysA,torus}). Such a new coordinate system is parametrized by:

\begin{equation}\label{tori}
x=\frac{a\sinh b\cos\varphi}{\cosh
b-\sin\eta},\hspace{1cm}y=\frac{a\sinh b\sin\varphi}{\cosh
b-\sin\eta},\hspace{1cm}z=\frac{a\cos\eta}{\cosh b-\sin\eta}\,,
\end{equation}
where $\varphi$ and $\eta$ vary from $0$ to $2\pi$, while the new constant
parameters $a$ and $b$ are both real and positive, what impose that the above parametrization is valid only for the ring torus, $R>r$. In terms of $r$ and $R$, we have:
\begin{equation}\label{tori1}
a=\sqrt{(R+r)(R-r)}\,,\hspace{1cm}\cosh b=\frac{R}{r}\,,
\end{equation}
which allow the interpretation of $a$ and $b$ as the {\em geometrical radius} and the {\em eccentric angle}, respectively \cite{torus}. Conversely, we have $R=a/\tanh b$ and $r=a/\sinh b $. In addition, it is easy to obtain the relations $\phi=\varphi$ and $\tan(\eta/2)=
\sqrt{(R+r)/(R-r)}\tan(\theta/2)$. Therefore, in this set of coordinates, the Hamiltonian (\ref{Htoro}) describing toroidally symmetric solutions, $\Theta=\Theta(\varphi)$ and $\Phi=\Phi(\eta)$, gets the form:
\begin{equation}
H=J\int^{2\pi}_{0}d\varphi \, \int^{2\pi}_{0}d\eta\left[\frac{( \partial_{\varphi}\Theta)^{2}}{\sinh
b}+\sinh b\sin^{2}\Theta(\partial_{\eta}\Phi)^{2}\right]\,,
\end{equation}
while eqs. (\ref{TH},\ref{PH}) read as: 
\begin{equation}\label{dif2}
\sin\Theta\cos\Theta(\partial_{\eta}\Phi)^{2}=\sinh^{-2}
b\,\partial^{2}_{\varphi}\Theta \qquad {\mbox{\rm and}} \qquad \partial^{2}_{\eta}\Phi=0.
\end{equation}
The latter equation has the simplest solution $\Phi(\eta)=\eta+\eta_{0}$, which after substitution in the first one, gives us the sine-Gordon equation (with $\xi=\varphi\sinh b$):

\begin{equation}\label{sine}
\partial^{2}_{\xi}\Theta=\sin\Theta\cos\Theta,
\end{equation}
whose the simplest solution reads \cite{SaxenaPhysA,Rajaraman} :
\begin{equation}
\Theta(\varphi)=2\arctan(\text{e}^{\varphi\sinh b})\,\label{solitontoro}
\end{equation}
with energy
\begin{equation}
E_{ring}=8\pi J \tanh(\pi\sinh b)=8\pi J\tanh \left(\pi\sqrt{\frac{R^2}{r^2} -1}\right) \le 8\pi J\,,
\end{equation}
which is in agreement with the saturated Bogomol'nyi inequality \cite{bogo}, 
$E_{soliton}=8\pi J|Q|$, once if we evaluate the solitonic charge, $Q=\frac{1}{4\pi}\int \sin\Theta d\Theta d\Phi$, for the solution above we exactly obtain $Q_{\rm ring}=\tanh(\pi\sinh b)$. Figure \ref{Q-ring} shows how the solitonic charge behaves with the torus size, $R$. Namely, note that only when $R\to\infty$ ($ \sinh b\to \infty$) this charge equals unity. At this limit, the soliton agrees with its counterpart lying in an infinite cylinder and represents a complete mapping from the spin sphere to the target manifold (the torus), a $\pi$-soliton, so corresponding to the first homotopy class of the second homotopy group of the mapping of the spin sphere to the (infinite) torus, say, $\pi_2(S^2\to T^1|_{R\to\infty})={\mathbf Z}$. However, for finite $R$, such a mapping is incomplete and no homotopy arguments can be used for classifying solution (\ref{solitontoro}) as a topological excitation. Indeed, in this case, we must take into account the topology of the geometrical support: although the genus prevents the complete mapping from the spin sphere onto the torus (so that the solution presents a fractional charge, $|Q|<1$), at the same time it also ensures topological stability, at principle, preventing the fractional soliton from decaying against the ground-state. In other words, now the soliton acquires a finite characteristic length, proportional to the genus size, which prevents its collapse and consequently its size from vanishing. Similar scenarios are provided by the annulus, the truncated cone \cite{SaxenaPRB2002}, and by the punctured pseudosphere \cite{nossapseudoPLA}.\\ \\

Once the torus is the topological product of two circles, we may wonder whether another solitonic solution depending on the polar angle ($\theta$ or $\eta$), $\Theta(\theta)$, should not also appear in this framework. This is indeed the case. If we reconsider Hamiltonian (\ref{Htoro}) with the conditions $\lambda=0$, $\Theta=\Theta(\theta)$ and $\Phi=\Phi(\phi)$, we get:
\begin{equation}
H_2= J\int_{-\pi}^{\pi} \,d\phi \int_0^{2\pi}\,d\theta \left[ \left(\frac{r}{R+r\sin\theta}  \right)\sin^2\Theta (\partial_\phi \Phi)^2 +\left( \frac{R+r\sin\theta}{r}  \right)(\partial_\theta \Theta)^2  \right]\,,\label{Htoro2}
\end{equation}
from what we obtain:
\begin{equation}
\partial^2_\phi \Phi=0 \qquad {\rm and}\qquad \left(\frac{r}{R+r\sin\theta}  \right) \sin\Theta\cos\Theta(\partial_\phi \Phi)^2=\cos\theta (\partial_\theta\Theta) +\left( \frac{R+r\sin\theta}{r}  \right)(\partial^2_\theta \Theta)\,.
\end{equation}
The first equation is readily solved giving $\Phi=\phi+\phi_0$, so that the latter one is simplified to:
\begin{equation}
\left(\frac{r}{R+r\sin\theta}  \right) \sin\Theta\cos\Theta=\cos\theta (\partial_\theta\Theta) +\left( \frac{R+r\sin\theta}{r}  \right)(\partial^2_\theta \Theta)\,.\label{sGgen}
\end{equation}
This expression is a generalized sine-Gordon equation, with non-constant coefficients. At principle, equation above admits solitonic solutions (possibly with fractional charge), but a closed and analytical solution is still lacking \cite{SaxenaPRB2002}. A further light into this issue may be shed if we take $R\to0$ above. At this limit, equation above gets the simple form:
\begin{equation}
\sin\Theta\cos\Theta=\sin\theta\cos\theta (\partial_\theta \Theta) +\sin^2\theta (\partial^2_\theta \Theta)\,,\label{SGgensimple}
\end{equation}
that is easily solved by:
\begin{equation}
\Theta(\theta)=\pm\theta \label{solitonesfera}\,,
\end{equation}
which is precisely the simplest solution we have for the spherical geometry \cite{nossaesferaPLA,SaxenaPRB1995}, whose energy reads $E_{\rm sphere}=8\pi J$, corresponding to the fundamental solitons with charges $\pm 1$. Therefore, we have seen that solitonic solutions lying on a torus generally present (or are expected to) fractional charges. These charges equal unity only at specific limits: for $R\to\infty$ we effectively get a soliton on an infinite cylinder, while for $R\to0$ the spherical counterpart is obtained.\\

\section{Vortex-like solutions on the torus}
Geometrically, a vortex with a net winding number $\kappa\neq0$, may be viewed as a set of spins rotating in a
closed circuit around a core (whose center is a singular point) or a topological obstruction, making impossible to change the configuration
to a perfectly aligned state without tampering
with spins at an arbitrary distance from the core. Thus, once a vortex-like configuration cannot be continuously deformed to the ground-state, it acquire the status of a topologically stable excitation. Vortices have been intensively studied for decades in a number of physical systems, like superfluids and superconductors, and they have been recently observed in several nanomagnets as remanent states of the net magnetization \cite{ShinjoWacho}. In addition, they have been proposed to take part (and even to be the key elements) in some magnetoelectronic mechanisms with potential applications to magnetic recording, processing, sensors, and so forth \cite{rahm} .\\

When we model a vortex as a continuum of spins we intend to describe only its outer region, once inside the core the analytical treatment is expected to give
only an estimate of its energy, shedding no light about its real
structure and spins arrangement, which require numeric/simulation
techniques. However, as we shall see, in the toroidal topology, a
natural cutoff for the vortex is provided by the genus whenever $R-r>0$, so that, in a ring torus the solution presents no core. On the other hand, in the case of a horn torus ($R=r$) a singular core takes place, while for $R<r$ (spindle torus) two singular points are verified. In both cases, the cores appear at the self-intercepting points (discussed earlier, see Fig. \ref{horn-and-spindle} and related text).\\

\begin{figure}[!htb]
\begin{center}
\includegraphics[width=7cm]{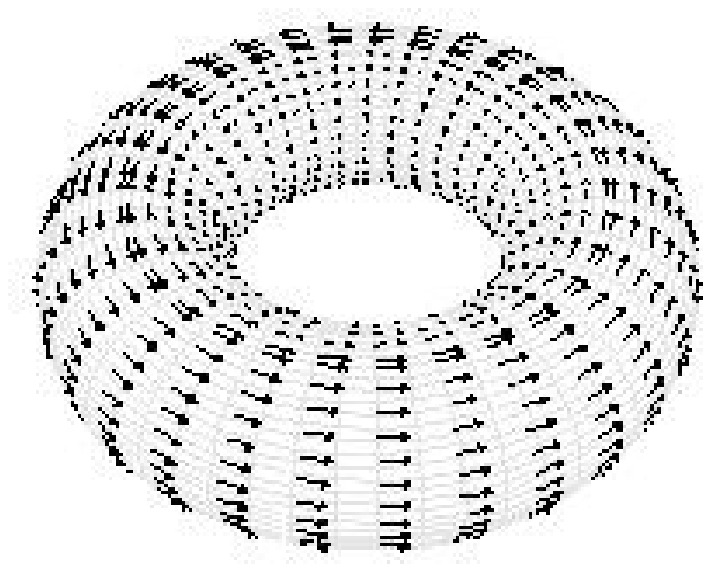}\hskip .5cm\includegraphics[width=5.5cm]{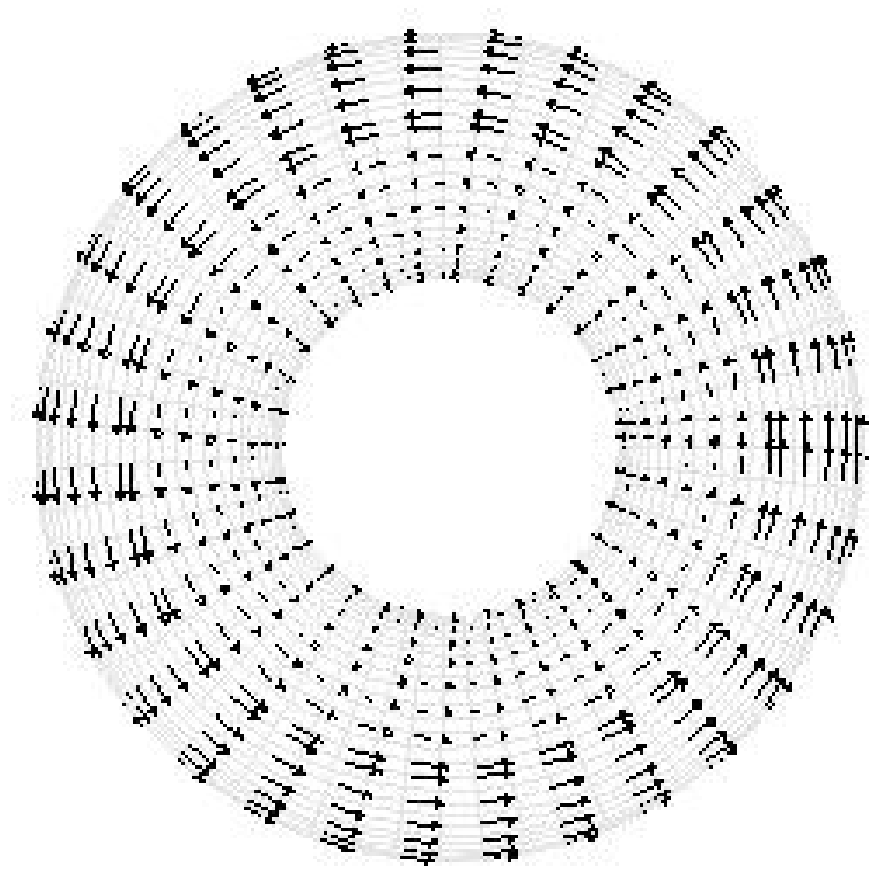}\caption{\small Global (left) and top (right) views of a vortex with
$\kappa=+1$ on the ring torus. The arrows represent the spin field, $\nabla\Phi$. The genus holds as a natural cutoff, preventing any core formation where energy could spuriously diverge.} \label{vorticetoro} \end{center}
\end{figure}

To achieve our purposes we shall consider the Planar Rotator Model, whose continuum Hamiltonian may be obtained from (\ref{Htoro}) with $\lambda=-1$ and
$\Theta=\pi/2$ (once we are dealing with static
solutions, our results apply equally well to the XY model), so that:
\begin{equation}
\label{Hprm} H=J\int_{0}^{2\pi}\int_{0}^{2\pi} \left[
\frac{r}{R+r\sin\theta}\left(\partial_{\phi}\Phi\right)^{2}+
\frac{R+r\sin\theta}{r}\left(\partial_{\theta}\Phi\right)^{2}
\right]d\phi d\theta\,.
\end{equation}
Demanding $\Phi$ to be cylindrically symmetric,
$\Phi=\Phi(\phi)$, Hamiltonian above yields the following:
\begin{equation}
\partial_{\phi}^{2}\Phi=0\qquad \Longrightarrow \qquad \Phi(\phi)=\kappa\phi+\phi_{0}\,,\label{solvortex}
\end{equation}
where $\kappa$ is the charge of the vortex while $\phi_{0}$ is a constant
accounting for its global profile, giving no contribution to its exchange
energy. The charge (vorticity) is formally defined, in the continuum
limit, as:
\begin{equation}
\kappa=\frac{1}{2\pi}\oint_{C}\left(\vec{\nabla}\Phi\right)\cdot
d\vec{l}\,,\label{kvortex}
\end{equation}
where the integration is evaluated along a closed path, C, around the
genus of the torus. Taking the solution above to the Hamiltonian
(\ref{Hprm}), we obtain, for a ring torus $(R>r)$:

\begin{equation}
E_{\rm v-ring}=4\pi^{2}J\kappa^{2}\frac{r}{\sqrt{R^{2}-r^{2}}}\,,\label{Evring1}
\end{equation}
from what we see that, distinctly from other cases (planar, conical, spherical or
pseudospherical geometries), the energy of a vortex in a ring torus does not present singularities, so that no cutoff needs to be introduced to
prevent spurious divergences. The profile of a vortex with $\kappa=+1$ and $\phi_0=0$ is presented in Fig. \ref{vorticetoro}. Note that its energy increases with $r$ and decreases as $R$ is raised. In addition, note that as $R\to\infty$ then $E_{\rm v-ring}$ vanishes, what is expected since at this limit we effectively deal with the spins lying along the axial direction of an infinite cylinder, so being parallel to each other (in the ferromagnetic case) . Thus, instead of a true vortex we have the ground-state configuration in this flat geometry, whose normalized energy vanishes.\\ 

Similar analysis may be suitably employed to the remaining supports. However, in the case of a horn torus ($R=r>0$), Hamiltonian (\ref{Hprm}) diverges for the vortex solution (\ref{solvortex}). This is not surprising since for this case the genus of the torus has been shrunk to a point (but it is not absent) where a singular core may be developed. Now, this spurious divergence must be bypassed by introducing a cutoff around the vortex core size, $l_0$. Doing that, the vortex energy on a horn torus may be easily evaluated to give:
\begin{equation}
E_{\rm v-horn}=4\pi J \kappa^2\, \cot(l_0/2r)\,,\label{Evhorn}
\end{equation}
which depends only upon the relative sizes of the core to the vortex, $l_0/r$. The singular core appears exactly where the horn torus is self-intercepting. This will also happen to the spindle torus, where a pair of singularities takes place, once this surface crosses itself at two distinct points (in variable $\theta$), like follows.\\

If we insist in configurations like $\Phi=\Phi(\phi)$, we clearly realize that the energy density diverges at those two points where the spindle torus is self-intercepting, say, $R+r\sin\theta=0$. Integrating Hamiltonian (\ref{Hprm}) with solution (\ref{solvortex}), but keeping $\theta$ arbitrary, gives:
\begin{equation}
\epsilon_{\rm spindle}= 2\pi \kappa^2 J\frac{r}{\sqrt{r^2-R^2}} \, \ln \left( \frac{r-\sqrt{r^2-R^2}+ R\tan(\theta/2)}{r+\sqrt{r^2-R^2}+ R\tan(\theta/2)} \right)\,. 
\end{equation}
For extracting some finite and meaningful quantity we must evaluate the result above over proper intervals. For that we should introduce suitable cutoffs around those points where energy density given by (\ref{Hprm}) blows up, say, $\theta_{\rm sing}=\arcsin(-R/r)$ and $\pi -\arcsin(-R/r)$, and only later we should evaluate $\epsilon_{\rm spindle}$ suitably in order to bypass the singularities. However, this task is very tedious and results in a very length expression, which will be omitted here. The main feature is that the well-defined vortex energy clearly shows the appearance of two singular cores located at both $\theta_{\rm sing}$, given above. Another interesting issue is that whenever $R\to0$, then integration of Hamiltonian (\ref{Hprm}) for solution (\ref{solvortex}) gives us exactly that result obtained in the spherical case\cite{nossaesferaPLA}, namely, exhibiting the pair of cores at antipodal points. In summary, if we start off by a ring torus ($R>r$) where the vortex is coreless and decrease $R$, we eventually get a horn torus, where a core is formed. Decreasing $R$ further a spindle torus is obtained and the vortex now presents a pair of singular cores, at the self-intercepting points. At the limit $R\to0$ these cores are located at diametrically opposite positions and, effectively we have a vortex lying on a sphere of radius $r$.\\

As previously discussed for the solitonic case, we could look for `vortex-like' solutions  depending on the polar angle, say, $\Phi=\Phi(\theta)$. For this case, Hamiltonian (\ref{Hprm}) yields:
\begin{equation}
\partial_\theta \Phi= \kappa'  \frac{r}{R+r\sin\theta}\,,\label{vortextheta} 
\end{equation}
while the topological charge is given by $k_\theta\equiv \frac{1}{2\pi} \oint (\nabla\Phi)\cdot d\vec{l}_\theta= \kappa'/\sqrt{R^2-r^2}$ (so, valid only for the ring torus). Therefore, the solution of equation above  may be written as: 
\begin{equation}
\Phi(\theta)= 2\kappa_\theta \arctan\left( \frac{r + R\tan(\theta/2)}{\sqrt{R^2-r^2}}\right) + \theta_0\,,
\end{equation}
whose energy is easily evaluated and gives:
\begin{equation}
E_{\rm \theta-ring}= 4\pi^2 \kappa^2_\theta J \frac{\sqrt{R^2-r^2}}{r}\,,
\end{equation}
which blows up as the ring torus becomes infinite, $R\to \infty$.\\

\section{Conclusions and Prospects}
We have studied the Heisenberg exchange model for classical spins defined on a toroidal geometry/topology. Solitonic and vortex-type configurations were described in some details, including their energy, profile and behavior at some limiting cases of the torus size and geometry.\\

We first considered the isotropic regime. There, only for the ring torus ($R>r$) solitonic solutions were analytically obtained. These solutions appear to bear fractional charges while their stability could be ensured, at principle, by the non-trivial topology of this torus, provided by the finite size of its central hole, $R-r>0$. As $R$ increases such a charge is raised but equals the unity only for $R\to\infty$ (in practice, a soliton in an infinite cylinder). On the other hand, as $R\to0$ a soliton lying on a sphere is described.\\

Now, taking the XY regime we have investigated vortex-like configurations in this support. In a ring torus the vortex exhibits no core and its energy density is finite everywhere. Its net energy vanishes as $R\to\infty$, once the spins are now practically parallel each other (the ferromagnetic ground-state). If $R=r$ (horn torus) then the vortex develops a singular core at the self-intercepting point, while for $R<r$ (spindle torus) a pair of such cores appear, so that as $R\to0$ they tend to be located at antipodal points and we recover the spherical case.\\

An interesting problem to be investigated is a small nanoring, say, with the geometry and topology of a ring torus. Actually, the expression ``{\em nanorings}'' frequently appears in the Nanomagnetic literature, including some possibilities for actual applications \cite{ShekaVavassori}. Such an object is generally fabricated by making a sufficiently large centered hole in a thin cylindrical nanodisk. Although it shares the torus topology its geometry does not, once at the disk and hole borders curvature changes abruptly. In addition, magnetostatic energies, which are very sensitive to the size and geometry of the magnet, are relevant in these cases, so that their evaluation in smoother supports seems to be more manageable, what justifies our prospect. Furthermore, our results could have some relevance for other branches were topological excitations and/or toroidal surfaces concern.\\ 
\vskip 1cm
\centerline{\Large Acknowledgements\\}
\vskip .5 cm
The authors thank CNPq and Fapemig for partial financial support.\\

\end{document}